\documentclass[sigconf,authorversion]{acmart}

\usepackage{tabularx}
\usepackage{booktabs}

\usepackage[utf8]{inputenc}
\usepackage{natbib}
\usepackage{hyperref}
\usepackage{microtype}
\usepackage{mathtools}
\usepackage{algorithmic}
\usepackage{textcomp}
\usepackage{xcolor}
\usepackage{multirow}

\usepackage{csquotes}
\usepackage{siunitx}

\newcommand{\p}[1]{POWER#1}
\newcommand{\occmmi}{OCC main memory interface}
\newcommand{\hwmon}{\texttt{hwmon}}
\newcommand{\powernvplugin}{\emph{IBM PowerNV Score-P Plugin}}

\ifx\GnuplotFromScratch\undefined
\else
  \usepackage[cleanup]{gnuplottex} %
  \usepackage{xparse}
  \usepackage{pgfplots}
  \DeclareUnicodeCharacter{2212}{−}
  \usepgfplotslibrary{groupplots,dateplot}
  \pgfplotsset{compat=newest}

  \ExplSyntaxOn
  \DeclareExpandableDocumentCommand{\convertlen}{ O{cm} m }
  {
  \dim_to_decimal_in_unit:nn { #2 } { 1 #1 } cm
  }
  \ExplSyntaxOff
\fi

\DeclareSIUnit\sample{Sa}

\ifx\giga\undefined
 \DeclareSIPrefix\giga{G}{9}
\fi
\ifx\byte\undefined
\DeclareSIUnit\byte{B}
\fi

\DeclarePairedDelimiter\abs{\lvert}{\rvert}

\setcopyright{acmcopyright}
\copyrightyear{2023} 
\acmYear{2023} 
\setcopyright{acmlicensed}\acmConference[ICPE '23]{Proceedings of the 2023 ACM/SPEC International Conference on Performance Engineering}{April 15--19, 2023}{Coimbra, Portugal}
\acmBooktitle{Proceedings of the 2023 ACM/SPEC International Conference on Performance Engineering (ICPE '23), April 15--19, 2023, Coimbra, Portugal}
\acmPrice{15.00}
\acmDOI{10.1145/3578244.3583729}
\acmISBN{979-8-4007-0068-2/23/04}

\begin{document}

\ifx\DraftModeOn\undefined

  \newcommand{\todoti}[1]{}
  \newcommand{\todomb}[1]{}
  \newcommand{\todors}[1]{}
  \newcommand{\todoag}[1]{}
  \newcommand{\tododh}[1]{}
  \newcommand{\todoht}[1]{}

  \newcommand{\figref}[1]{Figure~\ref{fig:#1}}
  \newcommand{\tabref}[1]{Table~\ref{tab:#1}}
  \newcommand{\secref}[1]{Section~\ref{sec:#1}}
  \newcommand{\lstref}[1]{Algorithm~\ref{alg:#1}}

  \newcommand{\papertarget}[2]{}

\else

  \newcommand{\todoti}[1]{\todo[color=yellow!60,inline,size=\small]{Thomas: #1}}
  \newcommand{\todomb}[1]{\todo[color=cyan!60,inline,size=\small]{Mario: #1}}
  \newcommand{\todors}[1]{\todo[color=green!60,inline,size=\small]{Robert: #1}}
  \newcommand{\todoag}[1]{\todo[color=orange!60,inline,size=\small]{Andreas: #1}}
  \newcommand{\tododh}[1]{\todo[color=red!60,inline,size=\small]{Daniel: #1}}
  \newcommand{\todoht}[1]{\todo[color=pink!60,inline,size=\small]{Hannes: #1}}

  \newcommand{\figref}[1]{\textcolor{red}{Figure~\ref{fig:#1}}}
  \newcommand{\tabref}[1]{\textcolor{red}{Table~\ref{tab:#1}}}
  \newcommand{\secref}[1]{\textcolor{red}{Section~\ref{sec:#1}}}
  \newcommand{\lstref}[1]{\textcolor{red}{Algorithm~\ref{alg:#1}}}

  \newcommand{\papertarget}[2]{\todo[inline]{Target: #1: \\Deadline: #2}}

\fi

\title{Evaluating the Energy Measurements of the IBM \p{9} On-Chip Controller}

\author{Hannes Tröpgen}
\orcid{0000-0001-9601-8683}
\affiliation{%
  \department{Center for Information Services and High Performance Computing (ZIH)}
  \institution{Technische Universität Dresden}
  \postcode{01062}
  \city{Dresden}
  \country{Germany}
}
\email{hannes.troepgen@tu-dresden.de}

\author{Mario Bielert}
\orcid{0000-0003-3363-1776}
\affiliation{%
  \department{Center for Information Services and High Performance Computing (ZIH)}
  \institution{Technische Universität Dresden}
  \postcode{01062}
  \city{Dresden}
  \country{Germany}
}
\email{mario.bielert@tu-dresden.de}

\author{Thomas Ilsche}
\orcid{0000-0002-5437-3887}
\affiliation{%
  \department{Center for Information Services and High Performance Computing (ZIH)}
  \institution{Technische Universität Dresden}
  \postcode{01062}
  \city{Dresden}
  \country{Germany}
}
\email{thomas.ilsche@tu-dresden.de}



\begin{abstract}
  Dependable power measurements are the backbone of energy-efficient computing systems.
  The IBM PowerNV platform offers such power measurements through an embedded PowerPC 405 processor: The \emph{On-Chip Controller} (OCC).
  Among other system-control tasks, the OCC provides power measurements for several domains, such as system, CPU, and GPU.
  This paper provides a detailed description and an in-depth evaluation of these OCC-provided power measurements.
  For that, we describe the provided interfaces themselves and experimentally verify their overhead (\SIrange{3.6}{10.8}{\micro\second} per access) and readout rate (\SI{24.95}{\sample\per\second}).
  We also study the consistency of the reported sensor readouts across the measurement domains and compare it to externally measured data.
  Furthermore, we estimate the internal sampling rate (\SI{1996}{\sample\per\second}) by provoking aliasing errors with artificial workloads,
  and quantify the errors that such aliasing could introduce in practice (for power consumption of processors \SI{12}{\percent} in our experimental worst-case scenario).
  Given these insights, practitioners using the IBM PowerNV platform can assess the quality of the embedded measurements,
  permitting sought-after energy efficiency improvements.
\end{abstract}


\maketitle

\section{Introduction}\label{sec:intro}
\begin{figure}
  \centering
  \includegraphics[width=\columnwidth]{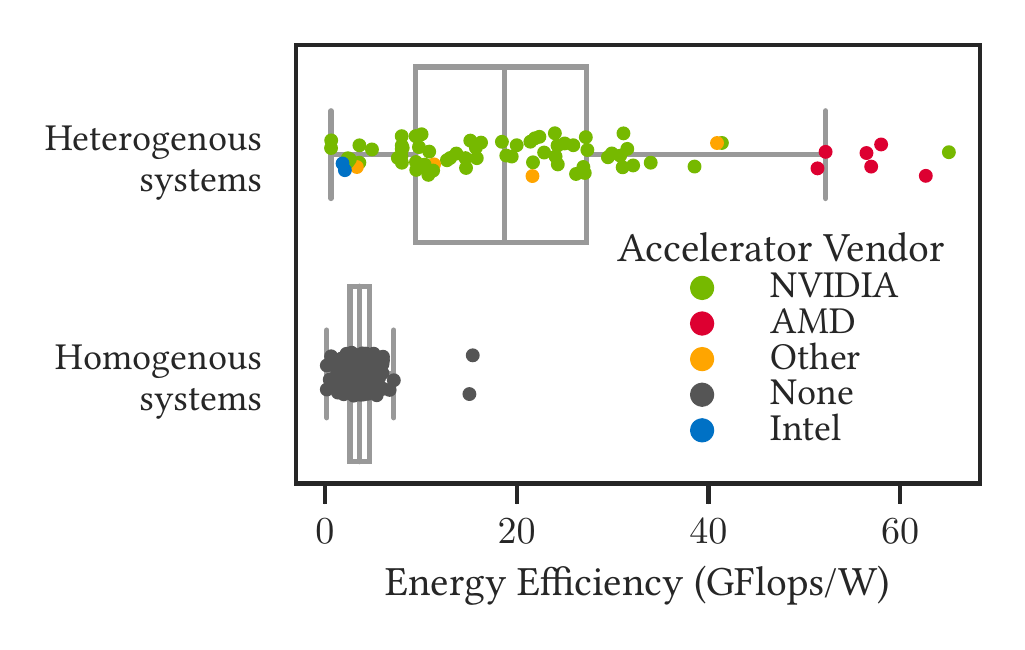}
  \caption{Energy efficiency of heterogenous and homogenous \emph{High Performance Computing} (HPC) Clusters according to the GREEN500 list compiled from \cite{Green500}.}\label{fig:green500_eff}
  \Description{Boxplots for energy efficiency in GFlops/W, split for homogenous and heterogenous HPC clusters. Homogenous clusters are clearly less energy efficient, their boxplot (including antennas) is below the box of heterogenous clusters.}
\end{figure}

The growing demand for accelerated computing, especially in machine learning and \emph{artificial intelligence} (AI), leads to the development of heterogenous architectures comprising prevalent multi-purpose processors and accelerators.
For example, processors of the IBM PowerNV platform bundle a high-core-count processor with several NVIDIA accelerators connected through NVLink.
IBM geared this architecture towards scalable and data-intensive workloads.
Even though such heterogenous systems are generally more energy-efficient (see \figref{green500_eff}) when used for suitable tasks, 
there are still parameters that influence the effective energy efficiency,
e.\,g., voltage/frequency selection~\cite{Mei_2013_measurementstudyGPU}.
Consequently, tuning such parameters to optimize energy efficiency depends on reliable power measurements.

This paper thoroughly evaluates the embedded power measurement interface on a \p{9} system as one representative of the PowerNV platform.
We describe the available measurement interfaces and their characteristics, such as readout latency and resolution.
We present the functional measurement domains and examine their accuracy.
Lastly, we use artificial workloads to examine the behavior of the internal measurement setup.
We want to establish how reliable the embedded power measurements are,
and ultimately allow for informed decisions regarding the energy efficiency of applications running on the PowerNV platform.

The remainder of this paper is structured as follows:
In the next \secref{relwork}, we introduce the background for power measurements and give an overview of the PowerNV processors.
Their power measurement interface itself is then described in \secref{occ},
and the interface's update rate and readout overhead are shown in \secref{interface_properties}.
After that, we focus on the measured values themselves and examine their accuracy in \secref{accuracy},
where we also demonstrate our measurement setup suitable for application tracing and profiling.
In \secref{samplerate_acc} we experimentally determine the internal sampling rate of the measurement.
The final \secref{summary} summarizes our work and sketches an outlook for future research directions.

Artifacts, including our raw results and programs used to obtain them, can be found online~\cite{artifacts}.\footnote{\url{https://github.com/tud-zih-energy/2023-power9-occ}}
This artifacts archive consists of multiple subdirectories, which are referred to throughout this paper with the prefix \path{artifacts/}, e.\,g.:

The data used for \figref{green500_eff} can be found in \path{artifacts/green500}.

\section{Related Work}\label{sec:relwork}

This section gives an overview of existing work.
First, we cover available measurements of the power consumption of computing systems in general.
Second, we describe prior work on the investigated architecture and its On-Chip Controller.

\subsection{Power Measurements of Computing Systems}

While dedicated precision power analyzers can provide excellent accuracies and sampling rates, their cost and space requirements are prohibitive for systems with multiple compute nodes.
However, several components in a modern data center power distribution provide power measurements out-of-the-box.
Some \emph{power distribution units} (PDUs) offer revenue-grade energy metering, e.g., Raritan PDUs with a \SI{1}{\percent} accuracy per ISO/IEC 62053-21~\cite{Raritan}.
In Section~\ref{sec:accuracy}, we use a monitored IBM PDU\footnote{Model number: 01KL833(46M4002)} that provides power readouts via the \emph{Simple Network Management Protocol} (SNMP).
To the best of our knowledge, there is no specification of the accuracy or quality of its power measurements.

Many server nodes also offer power monitoring via the \emph{Baseboard Management Controller} (BMC), typically measured at the \emph{power supply unit} (PSU).
The most prevalent protocols to read the power measurement data and other sensor data are the \emph{Intelligent Platform Management Interface} (IPMI)~\cite{IPMI} and the more recent Redfish~\cite{Redfish}.
The data can be read in-band or out-of-band via a network connection to the BMC and typical readout rates are \SI{1}{\sample\per\second} (\emph{1 sample per second}) or lower.
Such measurements are not always reliable, for example, Hackenberg et al.~\cite{Hackenberg_2013_ISPASS} have shown that Dell's implementation of power measurements via IPMI exhibits severe aliasing errors despite a documented \SI{1}{\percent} accuracy.

Moreover, many modern server processors provide power or energy measurements at the CPU level.
The most prominent example is the \emph{Running Average Power Limiting} (RAPL) mechanism originally developed by Intel~\cite{Rotem_2012_RAPL} but now also implemented by AMD.
Intel's documentation~\cite[Chapter~14.10]{intel_manual} describes that the RAPL registers provide an energy counter that is updated at $\sim$\SI{1}{\kilo\sample\per\second}.
However, Lipp et al.~\cite{Lipp_2021_PLATYPUSSoftwarebased} report higher update rates of up to \SI{20}{\kilo\sample\per\second} for certain domains.
Naively computing the average power consumption for short code regions may lead to inaccuracies since RAPL does not provide an update counter and thus the actual measurement duration is unknown.
Hähnel et al.~\cite{Haehnel_2012_RAPL_short} have demonstrated how to overcome this limitation and use RAPL to accurately measure regions in the order of a few milliseconds.
Several studies have shown that the quality highly depends on the specific implementation in the micro-architecture.
In particular, when the energy counter is implemented with a model rather than a physical measurement, the values can have biases towards certain workloads (see~\cite{Hackenberg_2013_ISPASS,rapl_dram,Hackenberg_2015_Haswell,Schoene_2019_SKL,Schoene_2021_Zen2}).

Several projects have demonstrated scalable approaches to add more sophisticated measurement infrastructure to compute clusters.
Examples are ArduPower~\cite{Dolz_2015_ArduPower}, PowerSensor~2~\cite{Romein_2018_PowerSensor2}, \cite{Ilsche_2018_HDEEM_HAEC}, and DiG~\cite{Libri_2019_DiG_OoB}.
The available readout rates for these solutions range from \SI{1}{\kilo\sample\per\second} to \SI{50}{\kilo\sample\per\second}, but some use higher internal sampling rates for increased accuracy.

\subsection{\p{8}, \p{9} \& The On-Chip Controller}\label{sec:p8_p9_occ}
In their intro to the \p{8} processor,
Fluhr et al.~\cite{p8_cpu_arch} place emphasis on the requirement to optimize energy and
consequently introduce the \emph{On-Chip Controller} (OCC) and sketch its monitoring and control capabilities.
The OCC is an embedded PowerPC 405 processor running a \emph{real-time operating system} (RTOS)~\cite[Sec. II]{p8_cpu_arch}.
The OCC's documentation states its main goals are to \textcquote[Sec. 1.2]{occ_p9_doc}{keep the system [thermally \& power] safe},
while leaving the P-state selection up to the \emph{operating system} (OS).
Additionally, the collected sensor data should be provided for external display~\cite[Sec. 1.2]{occ_p9_doc}.

The OCC persists across the PowerNV platform from its launch with \p{8} to the latest \p{10} processors.
However, the focused features shifted for the \p{9} processors:
Gonzalez et al.~\cite{p9_cpu_arch,p9_cpu_overview} describe it as a \textcquote[Sec. II]{p9_cpu_arch}{scale-out (SO)} processor,
highlighting its \emph{input/output} (IO) capabilities, namely a total \SI{300}{\giga\byte\per\second} accelerator, \SI{192}{\giga\byte\per\second} PCIeGen4 and \SI{230}{\giga\byte\per\second} memory bandwidth.
They also sketch the general architecture including the power domains,
thereby explaining some of the OCC's sensors.

The OCC itself runs a firmware of the same name,
available under the terms of the Apache 2 license~\cite{occ}.
Its architecture, interfaces, and capabilities are sketched in \secref{occ}.

Several submissions for OpenPOWER Summits cover the OCC,
including Rosedahl~\cite{occ_presi_overview} with a general introduction,
and Bhat~\cite{occ_presi} with a sketch of possible applications.

\section{Measuring Power Using the On-Chip Controller}\label{sec:occ}
The OCC manages the hardware sensors of the system and reports their readouts to the higher levels.
In total, it can report data for \num{75} types of sensors~\cite[Sec. 11.3]{occ_p9_doc}.\footnote{
  \num{31} of the \num{75} sensor types may only be collected out-of-band through \emph{Automated Measurement of Systems for Temperature and Energy Reporting} (AMESTER~\cite{amester})~\cite[Sec. 11.3.3]{occ_p9_doc}.
}
While the existence of an individual sensor is not guaranteed,
over \num{300} individual sensors were available during all of our tests.
These sensors report various measurement values,
including temperature, voltage, current, and frequency,
but also more abstract formats, e.\,g., processor \textcquote[Sec. 11.3.2.4]{occ_p9_doc}{utilization},
or event counts.
The scope varies from sensor to sensor,
most report on the core or processor level.
\tabref{sensors_power} lists all supported power sensors.

\begin{table}
  \centering
  \caption{OCC power sensors (without APSS), excerpt from~\cite[Sec. 11.3.2.2]{occ_p9_doc} with our domain descriptions}\label{tab:sensors_power}
  \begin{tabular}{l l r}
    Name & reported once per \ldots & Sampling interval (ms) \\
    \multicolumn{3}{l}{$\hookrightarrow$ Domain} \\ \midrule
    \texttt{PWRSYS} & system                   & 0.5                    \\
    \multicolumn{3}{l}{$\hookrightarrow$ bulk power consumption, measured after PSU output}\\
    \texttt{PWRGPU} & processor                & 0.5                    \\
    \multicolumn{3}{l}{$\hookrightarrow$ GPUs connected to this processor}\\
    \texttt{PWRMEM} & processor                & 0.5                    \\
    \multicolumn{3}{l}{$\hookrightarrow$ memory connected to this processor}\\
    \texttt{PWRPROC} & processor                & 0.5                    \\
    \multicolumn{3}{l}{$\hookrightarrow$ this processor itself (without attached components)}\\
    \texttt{PWRVDD} & processor                & 1.0                    \\
    \multicolumn{3}{l}{$\hookrightarrow$ processor cores, see $V_{dd}$ in \cite[Fig.~1]{p9_cpu_arch}}\\
    \texttt{PWRVDN} & processor                & 1.0                    \\
    \multicolumn{3}{l}{$\hookrightarrow$ processor nest, see $V_{dn}$ in \cite[Fig.~1]{p9_cpu_arch}}\\
  \end{tabular}
\end{table}

The OCC also reports up to \num{16} further power sensors from the \emph{Analog Power Subsystem Sweep} (APSS),
whose assignment is system-dependent and thus not considered in any of the experiments.
Some, but not all APSS-supported sensors are exposed by the OCC:
IO, storage, or fans power consumptions are only available through the APSS sensors~\cite[Sec. 6.3.1]{occ_p9_doc}.

The processor powers domains $V_{dd}$ (cores) and $V_{dn}$ (nest) are reported by the OCC,
although they do not fully cover the entire processor.
Besides the power of core and nest voltages,
the \p{9} processor distinguishes voltages for e.\,g., caches, memory, and other IO~\cite{p9_cpu_overview}.

On Linux systems, the OCC's sensor data is available through two in-band interfaces: \hwmon{} and the \occmmi{}.
Additionally, the \emph{OCC Poll Response} interface allows for individual sensors to be polled,
e.\,g., by the BMC~\cite[Sec. 2 \& 6]{occ_p9_doc}.
We did not consider the Poll Response interface in our experiments due to security considerations:
The Poll Response interface is only accessible for the BMC and \emph{Host Thermal Management} (HTMGT),
and requires granting write privileges for users to their respective memory regions for communication.
These same interfaces grant administrative permissions,
which is undesirable for regular users.
Granting access purely to the \occmmi{}/\hwmon{} is less intrusive, as read-only access is sufficient.

Overall the functionality is similar to those of classical BMCs via IPMI/Redfish.
The power measurements with six different kinds of measurement locations are comparatively sophisticated.

\subsection{The Linux Kernel Interface \hwmon{}}\label{hwmon}
\hwmon{}~\cite{hwmon} is the standard interface for \emph{hardware monitoring} of the Linux kernel.
Sensor data may be queried independently of the underlying hardware
by reading files under \path{/sys/class/hwmon} (\emph{sysfs}),
using the library \texttt{libsensors},
or the program \texttt{sensors}.
The reported values are provided by different drivers,
depending on the hardware~\cite[\texttt{Documentation/\allowbreak{}hwmon/\allowbreak{}hwmon-\allowbreak{}kernel-\allowbreak{}api\allowbreak{}.rst}, \texttt{Documentation/\allowbreak{}hwmon/\allowbreak{}sysfs-\allowbreak{}interface\allowbreak{}.rst}]{linux}.

The Linux kernel reads the list of available sensors and creates the corresponding \hwmon{} entries~\cite[\path{drivers/hwmon/ibmpowernv.c}]{linux}.
When accessing a sensor through \hwmon,
the corresponding callbacks read the requested value from the \occmmi{}~\cite[\path{hw/occ-sensor.c}, l.\,246\,ff.]{skiboot}.
Even though \hwmon{} indirectly uses the same data source as all other programs reading sensor data on PowerNV,
through its more generalized structure,
it can not represent all aspects of the values:
E.\,g., the OCC provides the sensor data in a buffer, which is updated regularly, consequently providing a timestamp of the last update~\cite[Sec. 11.3.1.3.1]{occ_p9_doc}.
\hwmon{} employs callbacks~\cite[\path{Documentation/hwmon/hwmon-kernel-api.rst}, l.\,147\,ff.]{linux}
and hence treats the values as being read on demand in real-time,
ignoring said timestamp under this assumption.

A patch series exposing more details from the OCC to \hwmon{} has been discussed on the Linux kernel mailing list~\cite{occ_linux_ml_hwmon_driver},
but was ultimately not implemented.

\subsection{The OCC Main Memory Interface}\label{sec:occmmi}
The OCC is connected to the main memory and periodically writes data for its sensors through this connection.
Note that the OCC documentation~\cite{occ_p9_doc, occ_p8_doc} does not use a consistent name for this interface,
calling the corresponding chapter \emph{OCC Main Memory Sensor Data}, however not using this term anywhere else.
We will use the term \emph{\occmmi{}} to refer to the interface described here.

The Linux kernel exposes the memory region to which the OCC writes sensor data to the userspace \emph{as-is} at \path{/sys/firmware/opal/exports/occ_inband_sensors}.~\cite[\path{arch/powerpc/platforms/powernv/opal.c}, l.\,881\,ff.]{linux}
Every OCC (i.\,e., every processor) creates one \emph{Sensor Data Block} of \SI{150}{\kilo\byte}
containing data for this particular processor.
Additionally, the first block also contains data for the entire system,
e.\,g.\ the bulk power consumption.
Every block contains two data buffers (\emph{ping} and \emph{pong buffer}),
which are used in an alternating fashion,
such that always at least one buffer is not being written to and hence contains valid data~\cite[Sec. 11.3]{occ_p9_doc}.

\begin{table}
  \centering
  \caption{Power sensor data reported by the \occmmi{}~\cite[Sec. 11.3.1.3]{occ_p9_doc}}\label{tab:occ_format}
  \begin{tabularx}{\columnwidth}{l c X}
    name                 & size   & description                             \\
                         & (byte) &                                         \\ \midrule
    \texttt{gsid}        & 2      & global sensor ID                        \\
    \texttt{timestamp}   & 8      & \SI{512}{\mega\hertz}-based timestamp  \\
    \texttt{sample}      & 2      & measured value                          \\
    \texttt{accumulator} & 8      & continuous sum                          \\
    \texttt{update\_tag} & 4      & number of samples stored in accumulator \\
  \end{tabularx}
\end{table}

The data reported for power sensors is shown in \tabref{occ_format}.
The resolution of an individual sample for all power sensors is \SI{1}{\watt}.
Notably, the OCC format~\cite[\path{src/occ_405/sensor/sensor_info.c}]{occ} would support a more fine-grained resolution---which is not used for any power sensor.
Even though the documentation promises an update rate of \num{1} or \SI{2}{\kilo\sample\per\second} for power sensors~\cite[Sec. 11.3.2.2]{occ_p9_doc},
the \occmmi{} update is only triggered every \SI{8}{\milli\second}~\cite[\path{src/occ_405/amec/amec_slave_smh.c}]{occ}.\footnote{
  We could not recreate these update rates in our experiments,
  neither the \SI{8}{\milli\second} \occmmi{} update rate (see \secref{samplerate_interface}),
  and \SI{2}{\kilo\sample\per\second} sample rate given in the documentation only with a measurable error (see \secref{samplerate_acc}).
}
Hence, the update rates of \num{1} or \SI{2}{\kilo\sample\per\second} only apply to the accumulator,
the exposed individual samples have a much lower update rate.
The \num{1} or \SI{2}{\kilo\sample\per\second} that make up this accumulator are not stored separately,
only their sum, i.\,e., the accumulator may be retrieved.
This leads to a theoretical resolution of \SI{1}{\milli\joule} (\SI{1}{\kilo\sample\per\second}) or \SI{0.5}{\milli\joule} (\SI{2}{\kilo\sample\per\second}) for the energy.
Due to this structure, in the following, we will distinguish between the \emph{external} update rate at which the interfaces expose their data,
and the \emph{internal} update rate at which the OCC operates, e.\,g.,
updates timestamps and the accumulator.

\section{Interface Properties}\label{sec:interface_properties}
In this section, we aim to measure the behavior of the interfaces themselves,
namely the readout latency and external update rate of \hwmon{} and the \occmmi{}.
The remaining system is not taken into account.

\subsection{Setup}
All experiments were performed on our local \emph{High Performance Computing} (HPC) cluster \emph{taurus}.

The jobs are distributed through the batch system and (exclusively) run on one of the 32 \p{9} nodes.
All 32 nodes are \emph{AC922} systems (code name \emph{Newell}, formerly \emph{Witherspoon}) by IBM.
Every node holds two \p{9} processors (model \texttt{02CY209}, code name \emph{Monza}) with 22 cores each,
resulting in a total of \num{176} threads with four-way \emph{simultaneous multithreading} (SMT) enabled.
Each processor has a \emph{thermal design power} (TDP) of \SI{250}{\watt};
the nominal frequency is \SI{2.8}{\giga\hertz} (up to \SI{3.1}{\giga\hertz} possible).
The machines use \SI{256}{\giga\byte} DDR4 main memory with a design bandwidth of \SI{170}{\giga\byte\per\second}.
They are primarily used for their six NVIDIA V100 (\emph{Volta}) GPUs, which are entirely ignored in these experiments.
The power is supplied by two \SI{2.2}{\kilo\watt} \emph{power supply units} (PSUs) following the standard 80+ Platinum~\cite{80plus}.
The PSUs run on \SI{230}{\volt} \emph{alternating current} (AC).
The nodes run \emph{Red Hat Enterprise Linux Server}, release 7.6 (\emph{Maipo}) as the operating system with a version 4.14.0 Linux kernel.
The OCC version is the commit \texttt{9047e57},
skiboot version \texttt{v6.5.3-29-g74a7a87a}.
The \occmmi{} is configured to be readable for non-root users.

This experiment's code and results are included in \path{artifacts/sampling_frequency_external_interface}.

\subsection{Approach}
To observe the external update rate and readout latency of the interfaces,
the system is used as-is, i.\,e.,
no specially crafted workloads are used.
The two interfaces (\hwmon{}, \occmmi{}) are observed separately, one after the other.

The measurement program collects $2^{24}$ samples in a loop using the respective interface and saves each read value together with a timestamp to an in-memory buffer.
After the execution is finished, this buffer is dumped into a file.
The program takes approx.~\SI{1}{\min} per run, chosen as a trade-off between accuracy and total experiment time.

For \hwmon{} the readout consists of a single read to the sysfs sensor file;
for the \occmmi{} the exposed sensor data is copied into a buffer,
from which the desired value is extracted following the specification~\cite[Sec. 11.3]{occ_p9_doc},
similar to an implementation presented by Bhat~\cite{occ_inband_demo}.
This extraction includes a lookup at which address the respective sensor is stored.
Although the specification does not guarantee that this sensor address stays constant~\cite[Sec. 11.3]{occ_p9_doc},
during test runs, it never changed.
Hence, we also include an optimized version to read the \occmmi{},
which breaks the specification by only reading the sensor address once and then only accessing the region for the respective sensor (instead of reading the entire file and always checking where the sensor data is located).
All methods allow us to monitor only one measurement domain at a time.
We use the bulk power of the system.%

We inspect the produced dumps for two separate properties:
First, we discuss the readout latency, i.\,e.,
the duration of a single access to the respective interface.
This ignores the reported data and purely uses the timestamps in the dumps.
In the second step, we examine the reported data itself to determine the \emph{external update rate}, i.\,e.,
at which interval new data is exposed by the OCC.

\subsection{Results}\label{sec:samplerate_interface}
\begin{figure}
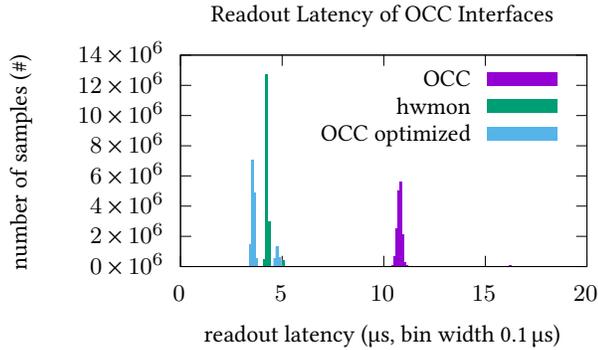

  \centering
  \ifx\GnuplotFromScratch\undefined
  \include{./img/paper-gnuplottex-fig1}
  \else
  \gnuplotloadfile[terminal=cairolatex, terminaloptions={size \convertlen{\columnwidth},\convertlen{0.6\columnwidth}}]{./img/sampling_frequency_interface/hist_readout_separation.gp}
  \fi
  \vspace{-0.7cm}
  \caption{Separation between two readouts of the OCC interfaces}\label{fig:sampling_overhead}
  \Description{Histogram of readout latency for different readout methods. There is almost no noise; see the text for a description of values.}
\end{figure}

The (mean) readout latency is the mean duration of a single interface access, i.\,e.,
the mean time between two successive interface readouts' timestamps.
The readout latency is \SI{4.3}{\micro\second} for \hwmon{}, \SI{10.8}{\micro\second} for the \occmmi{} (normal)/\SI{3.8}{\micro\second} (optimized) as \figref{sampling_overhead} details.
Both \hwmon{} and the normal \occmmi{} readout exhibit a single large spike in \figref{sampling_overhead}:
Their readout latency is practically constant.
(Tiny secondary spikes can be seen for both methods, which we consider negligibly small.
They are most likely caused by the scheduler briefly halting the execution of the experiment.)
This is not the case for the optimized access to the \occmmi{}.
Here, \figref{sampling_overhead} shows two spikes.
These originate in the two used data buffers (ping and pong buffer, see \secref{occmmi}).
If only one buffer contains valid data, the overhead is lower (\SI{3.6}{\micro\second} mean).
If both buffers contain valid data,
the readout routine checks which buffer has a newer timestamp.
This additional check increases the overhead to \SI{4.8}{\micro\second} (mean),
which was required for \SI{16}{\percent} of the samples in this experiment.

On average, %
the value provided by the interface changed every \SI{40.08}{\milli\second} for both the \occmmi{} and \hwmon{}:
The external update rate is \SI{24.95}{\sample\per\second}.
To compensate for a sensor reporting the same value in a new measurement interval again,
only value changes within a \SI{60}{\milli\second} window have been respected.
Due to its nature, the accumulator always changes between two readouts and is not affected by this.

The resolution of this experiment is limited by the duration of a single sample collection, i.\,e., the readout latency.
As this is three orders of magnitude smaller than the external update rate,
its determination remains accurate enough for the purposes of this experiment.

This difference in the orders of magnitude also makes the \SI{60}{\percent}\nobreakdash-gap between reading \hwmon{} vs.\ reading the \occmmi{} negligible:
One readout per update interval introduces approximately less than \SI{0.03}{\percent} overhead, considering a single thread.

As discussed in \secref{occmmi}, updates are expected exactly at multiples of \SI{8}{\milli\second}, i.\,e., here every \SI{40}{\milli\second}.
However, with \SI{40.08}{\milli\second} the measurement deviates from that by approx. \SI{0.2}{\percent}:
The same difference occurs when measuring the internal sampling rate of the OCC (see \secref{samplerate_acc}).

Li et al.~\cite[Sec. II, III-A]{understanding_occ} performed similar measurements on Google's Zaius platform.
They use \hwmon{} and report a duration of \SI{17}{\milli\second} to query the sensors.
They do not describe the process of obtaining these in detail,
hence the factor of $\sim$\num{1300} between their and our results can't be explained here.
Moreover,
their mentioned properties of the sensors do not align with our specifications at hand:
They mention \textquote{processor core data}\footnotemark{} as an example with an update rate of \SI{8}{\milli\second}.
In a reiteration of this experiment monitoring the power consumption of a single processor,
the external update rate remained at approx. \SI{40}{\milli\second}.
\footnotetext{Notably, according to the current OCC specification~\cite{occ_p9_doc}, the smallest reported power domain is the processor. Data for individual cores are not available.} %

\section{Consistency across different measurements}\label{sec:accuracy}

\begin{figure*}
  \centering
  \includegraphics[width=\textwidth]{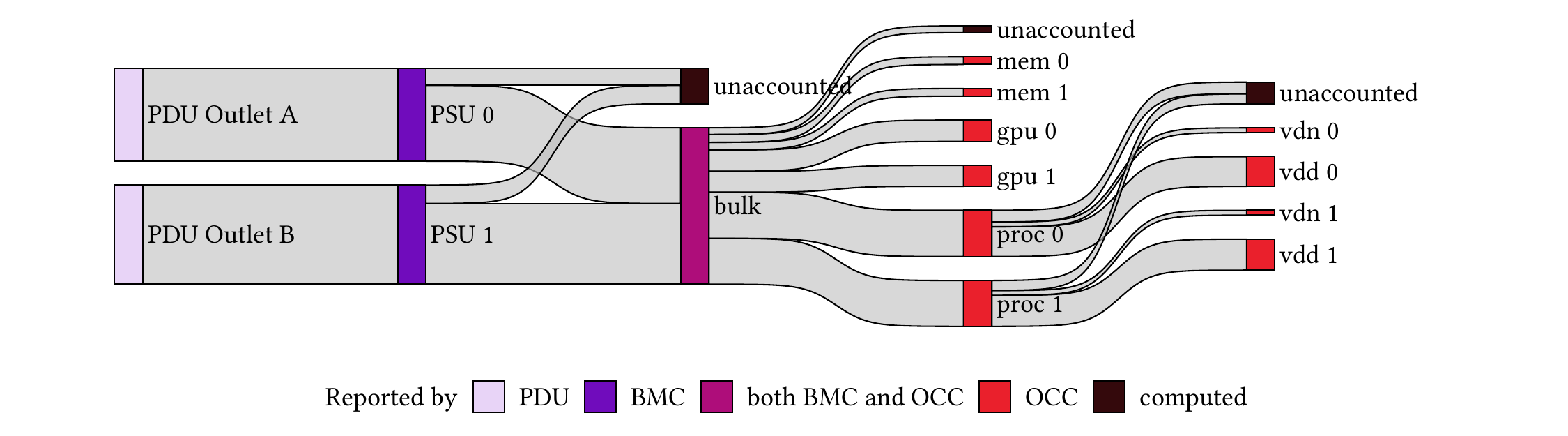}
  \caption{Power delivery scheme and monitored power domains. Colors indicate the data source. Band heights are scaled relatively based on a configuration from \secref{accuracy_approach}.}\label{fig:power_domains}
  \Description{Sankey plot of available power sensors. They show the following connections: 2 PDUs to 2 PSUs to unaccounted and system bulk, system bulk to unaccounted and OCC power sensors (twice) as listed above. Each Vdd and Vdn are sub-powers of each proc; additionally, proc subdivides into an accounted fraction. Only the bulk power is reported by OCC and BMC, all other sensors are only reported by a data source.}
\end{figure*}

To verify the values themselves, we compare them across different power measurement sources.
\figref{power_domains} shows the monitoring domains available to us along the power delivery path from the PDUs to the consumers.
First, the PDUs themselves report AC power output values via SNMP (cf. \secref{relwork}).
Next, the BMC reports both per-PSU input power as well as bulk power of the node.
The OCC also reports the bulk (total) power value.
Note that this value may not necessarily correspond to a single physical measurement point but could possibly be computed as a sum of multiple sensors at multiple voltages.
Finally, the OCC measurement points are detailed by \tabref{sensors_power}.

To quantify the accuracy, multiple data sources for the same measurement domain are necessary.
The only domain with more than one data source in our setup is the total/bulk power (reported by OCC and BMC).
In initial tests, we observed that the reported bulk power from the BMC matches the corresponding values from the OCC---but they can both plausibly come from the same sensors.
Therefore, we can not quantify the accuracy for any of the OCC's reported power domains.

Lacking multiple sensors for the same measurement domain,
we compare sensors of neighboring measurement domains:
(1)~OCC-reported bulk power against the PSU input, and
(2)~OCC-reported bulk power against the sum of individual components' power consumptions.
Since these comparisons do not cover the same measurement domain and even include a voltage conversion, a difference is expected.
That difference, as already shown in \figref{power_domains}, could contain both conversion losses as well as unaccounted consumers, e.g., small fans.

Nevertheless, this comparison demonstrates plausibility and would reveal workload biases.
Furthermore, the processor measurement domains comprise multiple voltages and also contain conversion losses and possibly unaccounted voltages, but we do not evaluate them in detail.
We confirmed that the PDU outlet and PSU input powers match closely.

\subsection{Measurement Setup}
For examination of the OCC's response to certain workloads,
the setup is much more similar to typical application profiling:
The desired workload is defined using the synthetic workload generator roco2~\cite{Hackenberg_2013_ISPASS},
which provides Score-P~\cite{scorep} user instrumentation.\footnote{
  When tracing an application, roco2 would be replaced by the instrumented application.
}
The OCC's values are loaded via the \powernvplugin{}, outlined in the following section.
The execution produces an OTF2 trace file~\cite{otf2},
which records OCC-reported values and the sections of the workload.
The raw trace and its processed forms are included in \path{artifacts/psu_comparison}.

\subsection{Score-P PowerNV Plugin}\label{sec:scorep_plugin}
To record the sensor readouts reported by the OCC, we developed a metric plugin for the Score-P plugin interface~\cite{scorep_plugin}, the \powernvplugin{}.\footnote{\url{https://github.com/score-p/scorep_plugin_ibmpowernv}}
This plugin reads the \occmmi{} and records all available power sensors at a configurable interval into an OTF2 trace.
For every OCC power sensor (see \tabref{sensors_power}) the plugin records the current sample, the timestamp, and the total energy based on the accumulator, as well as the number of samples in this accumulator.

According to the documentation~\cite[Sec. 11.3.2.2]{occ_p9_doc}, the accumulator has a sampling rate of \num{1} or \SI{2}{\kilo\sample\per\second},
depending on the sensor.
Therefore, between two readouts of the \occmmi{} (which are \SI{40}{\milli\second} apart), multiple samples are collected internally, i.e., \num{40} or \num{80} samples, respectively.
These samples are not exposed individually, but their sum is reported as the accumulator.
By tracing the changes in the number of accumulated samples and the accumulator itself,
the average power consumption can be computed:

\begin{displaymath}
  \label{eq:pwr_from_energy}
  power\ from\ energy(t_1, t_2) = \frac{accumulator(t_2) - accumulator(t_1)}{sample\ count(t_2) - sample\ count(t_1)}
\end{displaymath}

As this equation uses \emph{energy} (here the accumulator),
the result is referred to as \emph{power from energy}.

The resulting traces contain two metrics tracking the current power consumption in \si{\watt}:
The \emph{direct samples}, a single sample reported every \SI{40}{\milli\second} by the OCC,
and the \emph{power from energy}, the average power consumption of the last \SI{40}{\milli\second} based on \num{40} or \num{80} samples.

\subsection{Approach}\label{sec:accuracy_approach}

To gain a fine-grained profile of the sensor behavior, we test a wide range of power levels.
These levels are achieved by running seven workloads\footnotemark{} defined by roco2~\cite{Hackenberg_2013_ISPASS} on \num{1}, \num{2}, \ldots{}, \num{44} cores (with four threads per core).
During run-time, each workload executes for \SI{60}{\second} to create a stable environment, and to circumvent problems in the synchronization with the external data source.
This particular configuration is bundled as an example with roco2~\cite{Hackenberg_2013_ISPASS} under the name \emph{P9 Longrun}.
During execution, the power consumption as reported by the two PSUs was continuously collected and stored in the trace.
\footnotetext{
  These workloads are \texttt{busy wait}, \texttt{compute}, \texttt{matmul}, \texttt{memory copy}, \texttt{memory read}, \texttt{memory write}, and \texttt{sine}.
}

The tested system has two \SI{2.2}{\kilo\watt} power supplies,
whose power budget is not exhausted even with both processors under full load,
where the bulk power consumption is approximately \SI{1}{\kilo\watt}.\footnote{
  Their larger design capacity is due to the six \emph{graphics processing units} (GPUs) of the node,
  which are entirely ignored for this test.
}
One such configuration is shown in \figref{power_domains},
here the kernel \texttt{memory write} running on all cores draws a total \SI{1055}{\watt} for both PSU inputs combined.

\subsection{Results}
\begin{figure}
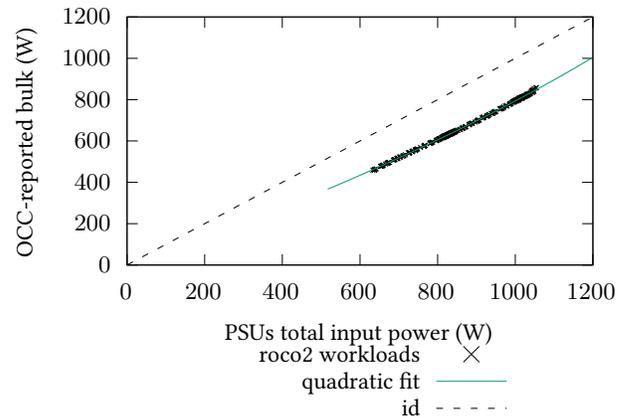

  \centering
  \ifx\GnuplotFromScratch\undefined
  \include{./img/paper-gnuplottex-fig2}
  \else
  \gnuplotloadfile[terminal=cairolatex, terminaloptions={size \convertlen{\columnwidth},\convertlen{0.7\columnwidth}}]{./img/psu_comparison/input_vs_occ_reported_sum.gp}
  \fi
  \vspace{-0.7cm}
  \caption{PSU input power vs.\ bulk power reported by OCC}\label{fig:psu_efficiency_reported}
  \Description{Scatterplot with bulk power and PSU input power on either axis. They show an almost linear relationship, only minimal noise is visible.}
\end{figure}

\figref{psu_efficiency_reported} compares the power consumption reported by the PSUs to the power consumption reported by the OCC.
Modeling the PSUs' efficiency using a quadratic fit\footnote{
  From our experience, PSUs do not exhibit linear efficiency.
  In this particular value range a linear regression would be sufficient,
  but still has approx. twice the error with \SI{0.4}{\percent} MAPE, \SI{3.2}{\watt} MAE.
} yields plausible results:
To this regression the workloads have a \SI{0.2}{\percent} \emph{mean absolute percentage error} (MAPE) and \SI{1.7}{\watt} \emph{mean absolute error} (MAE).
Across all workloads, the efficiency is \SI{77}{\percent}.
This low efficiency is rooted in the load on the PSUs with \SIrange{458}{859}{\watt} OCC-reported bulk power,
which is well below the design capacity of the two \SI{2.2}{\kilo\watt} PSUs.

\begin{figure}
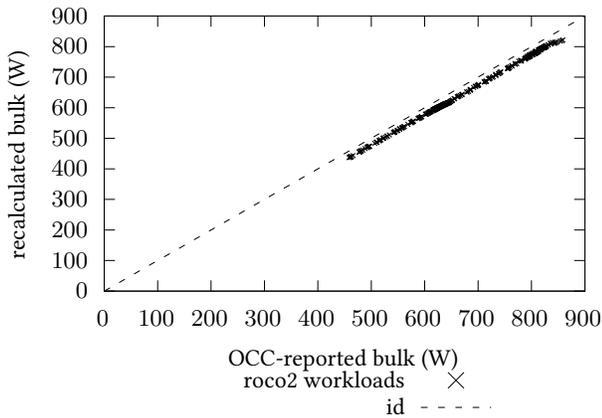

  \centering
  \ifx\GnuplotFromScratch\undefined
  \include{./img/paper-gnuplottex-fig3}
  \else
  \gnuplotloadfile[terminal=cairolatex, terminaloptions={size \convertlen{\columnwidth},\convertlen{0.7\columnwidth}}]{./img/occ_sums/occ_sums.gp}
  \fi
  \vspace{-0.7cm}
  \caption{Bulk power as reported by OCC vs.\ re-calculated sum}\label{fig:occ_bulk_vs_sum}
  \Description{Scatterplot with bulk power and re-calculated sum on either axis. The measurements visibly diverge from the expected identity, even though only slightly.}
\end{figure}

This experiment also uncovered a discrepancy in the OCC-re\-por\-ted bulk power:
Recalculating the bulk power consumption by adding the reported power consumptions of GPUs, CPUs, and memories has an inconsistent difference when compared to the reported bulk power of the system.\footnote{
  This is marked as \emph{unaccounted} in~\figref{power_domains}.
}
This discrepancy is shown in \figref{occ_bulk_vs_sum},
where the re-calculated sum vs.\ the reported bulk power consumption show \SI{3.8}{\percent} MAPE, \SI{25.5}{\watt} MAE.
The cause of this is not clear;
one or multiple components not measured individually, but included in the bulk power could be responsible for this difference.

The bulk powers reported by the OCC and the BMC match very well (MAPE \SI{0.2}{\percent}, MAE \SI{1.3}{\watt}).
The OCC and BMC could use the same underlying data source,
as the BMC could query the OCC using its Poll Response interface~\cite[Sec. 1.6]{occ_p9_doc}.
Further investigation of this possibility is not possible with the used setup.

Even though OCC and BMC-reported values match,
this experiment is no verification of the OCC-reported values:
We conclude that the OCC-reported values are plausible,
but may not verify their correctness.

\section{Internal Sampling Rate}\label{sec:samplerate_acc}
The OCC provides an accumulator for its power sensors but does not expose individual samples that contribute to its value.
In this section, we measure the internal sampling rate of the accumulator.

\subsection{Setup \& Approach}\label{sec:samplerate_acc_approach}
The setup is identical to \secref{interface_properties},
but lacking the PSU observation.
For simplicity, here we use a single sensor:
The power consumption of the first processor (\emph{proc 0}).\footnote{The approach can be applied to other sensors as long as a workload can be generated.}

\begin{figure}
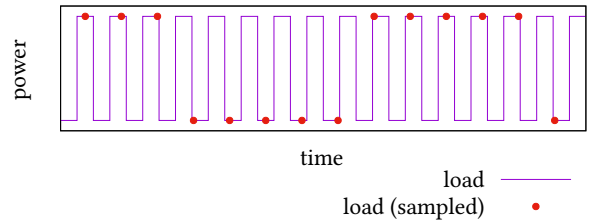

  \centering
  \ifx\GnuplotFromScratch\undefined
  \include{./img/paper-gnuplottex-fig4}
  \else
  \gnuplotloadfile[terminal=cairolatex, terminaloptions={size \convertlen{\columnwidth},\convertlen{0.4\columnwidth}}]{./img/aliasing/synthetisch.gp}
  \fi
  \vspace{-0.7cm}
  \caption{Concept of aliasing during sampling}\label{fig:aliasing_synth}
  \Description{Square signal to demonstrate aliasing. Sampled points are highlighted, and there is only approx. 1 sample per period. Consequently, the sampled points are not sufficient to reconstruct the square signal.}
\end{figure}

We assume that the OCC internally samples with \SI{2}{\kilo\sample\per\second}
(as the corresponding \textquote{sample time} is given with \SI{500}{\micro\second} in the specification~\cite[Sec. 11.3.2.2]{occ_p9_doc}),
and that these samples are added without any further processing into the accumulator.
Based on these assumptions,
we design a workload that idles when it is being sampled and creates a high load when it is not sampled (or vice versa),
effectively provoking aliasing.
For this, the frequencies of sampling and workload have to match perfectly---which, in practice, they do not:
A slight mismatch of frequencies causes this effect to shift over time, i.\,e.,
first only idle is sampled, after some time this shifts and only high load (work) is sampled, shifting back after some more time etc.
However, throughout this shift, there remains only approx. 1 sample per period.
This is sketched in \figref{aliasing_synth}.

This aliasing affects the traces indirectly:
Due to the update rate of the interface (approx. \SI{40}{\milli\second}),
the OCC does not expose all of these individual samples.
Only the mean of all accumulated samples since the last interface update is stored as \emph{power from energy} (see \secref{scorep_plugin});
here the mean across the last \num{80} samples is recorded to the trace.

Typically, some of the \num{80} samples are captured during the low power level,
and some during the high power level,
resulting in the power from energy-values hovering in between these two levels.
In particular,
as half the time is spent idle and half working,
the power from energy-values are the mean of the low and high power levels.
Hence, the sketch in \figref{aliasing_synth} would produce a stable power from energy level,
as every 5 samples the sampled power level changes between high and low level.

The power from energy deviates from the mean only if almost all of the 80 samples between two interface readouts are captured during idle (or almost all during work).
I.\,e., the aliasing only becomes apparent in the traces when workload and sampling rate are almost equal.

Combining this deviation from the mean of power from energy with the shift from sampling only high to sampling only low---due to small frequency deviations described above (see \figref{aliasing_synth})---yields typical aliasing patterns:
The power from energy slowly alternates between a higher and a lower level.
If we observe such a pattern,
we know the sampling rate and workload frequency are almost equal.
Based on the frequency of power from energy alternating between the two levels,
and the relation of this alternating pattern to the (known) frequency of the underlying workload changing between idle and work,
we can derive the sampling rate of the accumulator.

To produce such an effect,
we employed an alternating synthetic workload around \num{0.5}, \num{1}, \num{2}, and \SI{2.05}{\kilo\hertz}.
Half of each period is spent idling,
the other with computing.\footnote{
  The commands are selected to create low and high power levels.
  \emph{Idle} corresponds to setting medium thread priority~\cite[Sec. 3.2, p.\,838]{power_isa},
  high load is created by a naive vector dot product implementation.
  Setting the thread priority is also used by the Linux kernel to idle on PowerNV~\cite[\path{arch/powerpc/kernel/idle.c}]{linux}.
}
This configuration is bundled as an example with roco2~\cite{Hackenberg_2013_ISPASS} under the name \emph{P9 Highlow}.
The resulting traces and scripts for post-processing are provided in \path{artifacts/sampling_frequency_internal_accumulator}.

This particular workload creates another indicator for the aforementioned aliasing:
The \occmmi{} provides the latest individual sample (\emph{direct sample}),
which have a larger spread compared to the power from energy:
As the power from energy is based on an average across 80 samples,
which contains samples during idle \emph{and} work,
they are evening out---opposed to the single sample directly provided by OCC,
which is captured during \emph{either} idle \emph{or} work\footnotemark{}.
\footnotetext{or during the transition between those, but for this approach we consider this transition period to be negligibly short}
In a non-aliasing scenario, all direct samples exhibit a larger spread compared to the power from energy.

During aliasing,
all samples within the 80-sample window that power from energy uses are captured during idle (or all during work).
Consequently,
as all 80 samples record the same power level,
the mean is identical to the single direct sample:
When aliasing occurs,
the spread of power from energy and direct samples is pretty much identical.

\subsection{Results}
\begin{figure}
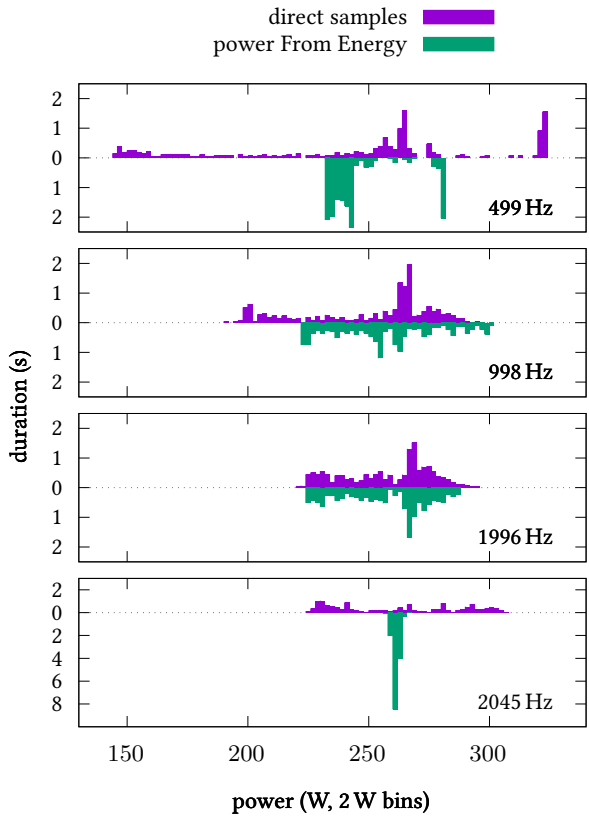

  \centering
  \ifx\GnuplotFromScratch\undefined
  \include{./img/paper-gnuplottex-fig5}
  \else
  \gnuplotloadfile[terminal=cairolatex, terminaloptions={size \convertlen{\columnwidth},\convertlen{1.3\columnwidth}}]{./img/sampling_frequency_accumulator/hist_acc_vs_sensor.gp}
  \fi
  \vspace{-0.7cm}
  \caption{Spread of direct samples versus power from energy for workloads of different frequencies}\label{fig:aliasing_spread}
  \Description{Histograms comparing the spread of direct samples and power from energy for frequencies 499 Hz, 998 Hz, 1996 Hz, and 2045 Hz. Only for 1996 Hz do direct samples and power from energy align; for all other frequencies they are different.}
\end{figure}

Exactly this effect on the spread of power from energy and direct samples can be observed:
For a workload with \num{499}, \num{998}, and \SI{2045}{\hertz} the spread of direct samples and power from energy values show a clear difference---which is lacking for a workload at \SI{1996}{\hertz}, as \figref{aliasing_spread} shows.

\begin{figure}
  \centering
  \scriptsize
  \includegraphics[width=\columnwidth]{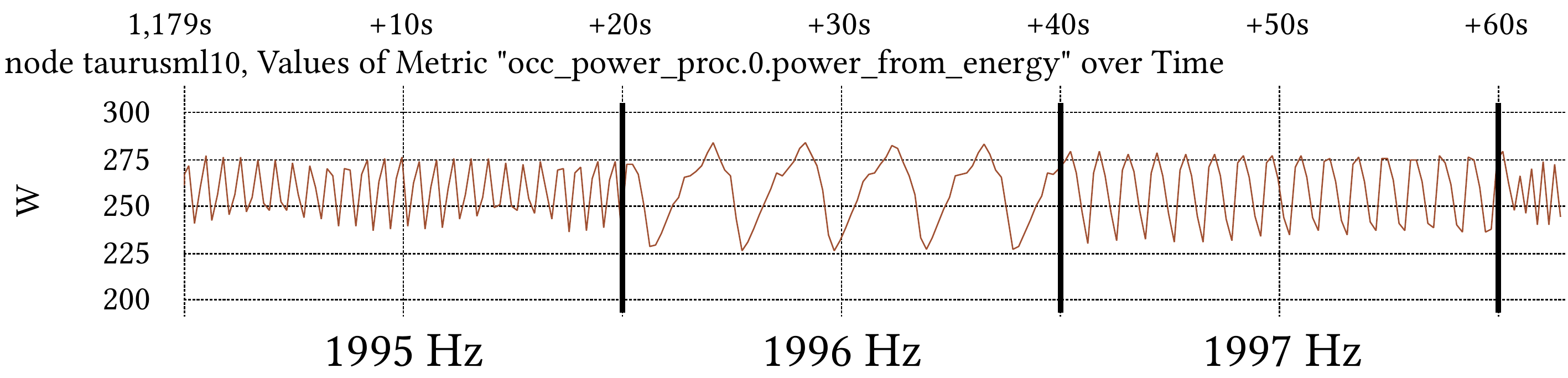}
  \caption{Power from energy for the first processor during workloads alternating with \SIrange{1995}{1997}{\hertz}}\label{fig:aliasing_pattern}
  \Description{Trace of power from energy. For each section with a constant frequency the power from energy slowly and regularly alternates between a high and a low level.}
\end{figure}

This indicates that aliasing is occurring for the \SI{1996}{\hertz} workload and only one sample per period (of the workload) is collected, as discussed above.
The aliasing pattern itself, 
i.\,e., the power from energy shifting between a high and a low level is shown in \figref{aliasing_pattern}.
(Note that the apparent slow shift between high and low power levels is purely a measuring artifact: The underlying workload still loops with \SIrange{1995}{1997}{\hertz}.)

\begin{table}
  \centering
  \caption{Internal OCC sampling rate, computed from the results in \figref{aliasing_pattern}}\label{tab:sampling_frequency}
  \begin{tabular}{r r r r}
    \multicolumn{2}{r}{$f_{pattern}$ (\si{\hertz})} & $f_{workload}$ (\si{\hertz}) & $f_{sampling}$ (\si{\hertz})                 \\ \midrule
    $24 \div \SI{19.4}{\second} \approx{}$            & \num{1.24}                     & \num{1995}                     & \num{1996.24} \\
    $4 \div \SI{16.5}{\second} \approx{}$             & \num{0.24}                     & \num{1996}                     & \num{1996.24} \\
    $14 \div \SI{18.3}{\second} \approx{}$            & \num{0.77}                     & \num{1997}                     & \num{1996.23} \\
  \end{tabular}
\end{table}

We use the pattern visible in \figref{aliasing_pattern} to compute the accumulator sampling rate.
For that, we manually read the frequencies of the emerging aliasing pattern from \figref{aliasing_pattern} and apply $\abs{f_{sampling} - f_{workload}} = f_{pattern}$,
which yields the results shown in \tabref{sampling_frequency}.
The internal sampling rate is approx. \SI{1996}{\sample\per\second},
which is \SI{0.2}{\percent} slower than the \SI{2}{\kilo\sample\per\second} noted in the specification~\cite[Sec. 11.3.2.2]{occ_p9_doc}---the same \SI{0.2}{\percent} as for the update rate of the interface noted above (see \secref{samplerate_interface}).

This leads us to hypothesize that the clocks of the system and the OCC diverge.
The experiment also recorded the timestamps and number of accumulated samples reported by the OCC:
Over the approx. \SI{30}{\minute} total experiment duration the clock deviates by less than one \emph{part per million} (PPM) from the expected \SI{512}{\mega\hertz},\footnote{
  The trace is recorded on the same system;
  but the accuracy of neither the trace nor the system is validated.
  For this particular experiment, a single update cycle of the interface takes \SI{40}{\milli\second},
  which corresponds to 20 PPM.
}
whereas the accumulator reports \num{1996.16} collected samples per second.

\subsection{Potential Errors and Implications}
For this experiment,
the power from energy slowly shifts between \SI{225}{\watt} and \SI{285}{\watt} as visible in \figref{aliasing_pattern}.
By further adjusting the frequency of the workload closer to the internal sampling rate,
it would be possible to stretch out this pattern further.
This may potentially hide its periodic behavior---in such a worst-case scenario only the low or only the high level might be sampled.
Assuming a \SI{255}{\watt} true average power consumption in the middle between both levels
(as the workload is split evenly between idle and work),
a sampling of only low (or only high) levels would yield a \SI{12}{\percent} error in comparison.
Such an error occurs only under very specific circumstances and represents a worst-case scenario.
Hence, for practical applications the error is likely much lower.

One should prefer the accumulator-based power from energy to measure power consumption,
as its sampling rate of \SI{\sim2}{\kilo\sample\per\second} yields more precise results compared to the individual samples reported with \SI{\sim25}{\sample\per\second} at every interface update (i.\,e., every \SI{\sim40}{\milli\second}).
The magnitude of the observed error depends on the ability of the monitored component to change the power consumption within one sampling period.
Observing GPUs or the entire system may result in different errors, depending on the particular setup.

In general, the observed \SI{0.2}{\percent} deviation from the advertised \SI{2}{\kilo\sample\per\second} internal sampling rate is hardly relevant to measuring applications:
As the OCC reports the number of accumulated samples (see \tabref{occ_format}),
the computation of the power from energy (see \secref{scorep_plugin}) remains correct.
For practical applications in general we do not expect a relevant impact in accuracy introduced through aliasing.

\section{Summary and Future Work}\label{sec:summary}
In this paper, we presented a detailed description of the \emph{On-Chip Controller} (OCC) of PowerNV platform processors.
We described the available power sensors (see \tabref{sensors_power}) and the interfaces that can be used to retrieve their readouts.
We measured that such readouts take between \SIrange{3.8}{10.8}{\micro\second} (mean) depending on the interface,
and that new values are provided every \SI{40.08}{\milli\second} (\SI{24.95}{\sample\per\second})---which is \SI{0.2}{\percent} slower than we expected.
Then, we compared the OCC measurement with an external measurement and presented our Score-P plugin to integrate these sensor readouts into OTF2 traces.
This comparison of values collected from PSUs, BMC, and OCC did not verify the correctness of the OCC-reported data,
but still confirms its plausibility.
Furthermore, we discovered an emerging aliasing effect for workloads with a frequency matching the OCC's internal sampling rate,
through which we produce \SI{12}{\percent} error in carefully crafted experiments.
This aliasing also exposes that the internal sampling rate is \SI{1996}{\sample\per\second},
which is \SI{0.2}{\percent} slower than specified in the documentation---similar to the external update rate of the interfaces.

The used workload generator only supports CPUs, hence the sensor behavior under load is only experimentally verified for CPUs.
By extending the workload generator to GPUs these experiments could cover more load scenarios.
All sensors, including those for GPUs, are processed by the same mechanisms of the OCC.
Hence, we expect no discrepancies to the general behavior described in the paper.
In particular, the principle for provoking the \SI{12}{\percent} error of the measured energy for CPUs through aliasing can be applied to GPUs as well,
but the magnitude of this error heavily depends on the installed hardware.
To confirm the values reported by the OCC additional sensors are required.
This can be achieved by manually instrumenting the node to measure the power consumption of the entire system, CPUs, and GPUs externally.

While in this paper we discussed the entire measurement pipeline of the OCC and quantified worst-case errors,
additional measurements, e.\,g., using external hardware, could further strengthen the confidence in the out-of-the-box power measurements provided by the OCC.

\begin{acks}
This work is supported in part by the German National High Performance Computing (NHR@TUD).
The authors are grateful to the Center for Information Services and High Performance Computing at TU Dresden for providing the Power9 Systems used in the measurements and the support during them.
\end{acks}


\end{document}